\documentclass{PoS}

\usepackage{amsmath,amssymb}
\usepackage{dsfont}
\usepackage{slashed}
\usepackage{verbatim}    
\usepackage{graphicx} 
 
\newcommand{\be}{\begin{equation}}
\newcommand{\ee}{\end{equation}}
\newcommand{\bea}{\begin{eqnarray}} 
\newcommand{\eea}{\end{eqnarray}}

\newcommand{\la}{\lambda}

\newcommand{\MSbar}{{\overline{\rm MS}}}

\def\openone{\leavevmode\hbox{\small1\kern-6.8pt\normalsize1}}

\def\lsim{\mathrel{\rlap{\lower4pt\hbox{\hskip1pt$\sim$}}
    \raise1pt\hbox{$<$}}}                
\def\gsim{\mathrel{\rlap{\lower4pt\hbox{\hskip1pt$\sim$}}
    \raise1pt\hbox{$>$}}}                

\newcommand{\Dlr}{\buildrel \leftrightarrow \over D\raise-1pt\hbox{}}
\newcommand{\Dl}{\buildrel \leftarrow \over D\raise-1pt\hbox{}}
\newcommand{\Dr}{\buildrel \rightarrow \over D\raise-1pt\hbox{}}

\title{One-loop lattice study of composite bilinear operators in Supersymmetric QCD}

\ShortTitle{One-loop lattice study of composite bilinear operators in SQCD}

\author{\speaker{M.~Costa}, H.~Panagopoulos\\
\llap{}Department of Physics, University of Cyprus, Nicosia, CY-1678, Cyprus\\
{\rm E-mail}: \email{kosta.marios@ucy.ac.cy}, \email{panagopoulos.haris@ucy.ac.cy}}

\abstract{We study 4-dimensional SQCD with gauge group $SU(N_c)$ and $N_f$ flavors of chiral super-multiplets on the lattice. We perform extensive calculations of matrix elements and renormalization factors of composite operators in Perturbation Theory. In particular, we compute the renormalization factors of quark and squark bilinears, as well as their mixing at the quantum level with gluino and gluon bilinear operators. From these results we construct correctly renormalized composite operators, which are free of mixing effects and may be employed in non-perturbative studies of Supersymmetry. All our calculations have been performed with massive matter fields, in order to regulate the infrared singularities which are inherent in renormalizing squark bilinears. Furthermore, the quark and squark propagators are computed in momentum space with nonzero masses.

This work is a feasibility study for lattice computations relevant to a number of observables, such as spectra and distribution functions of hadrons, but in the context of supersymmetric QCD, as a forerunner to lattice investigations of SUSY extensions of the Standard Model.
}

\FullConference{XIII Quark Confinement and the Hadron Spectrum - Confinement2018\\
31 July - 6 August 2018\\
Maynooth University, Ireland}
           
\begin{document}

\section{Introduction}
Current intensive searches for Physics Beyond the Standard Model (BSM) are becoming a very timely endeavor \cite{Tanabashi:2018}, given the precision experiments at the Large Hadron Collider and elsewhere; at the same time, numerical studies of BSM Physics are more viable due to the advent of lattice formulations which preserve chiral symmetry \cite{Luscher:1998}. In particular, the study of supersymmetric models on the lattice \cite{Curci:1986sm, Creutz:2001, Feo:2003, Giet&Poppitz, Suzuki&Tani:2005} has been an object of intense research activity in recent years \cite{Kaplan:2009, Catterall:2014vga, Joseph:2015xwa, Ali:2018fbq, Endrodi:2018ikq}, and applications to supersymmetric extensions of the Standard Model are gradually becoming within reach. Studies of hadronic properties using the lattice formulation of Supersymmetric Quantum Chromodynamics (SQCD) rely on the computation of matrix elements and correlation functions of composite operators, made out of quark ($\psi$), gluino ($\lambda$), gluon ($u$), squark ($A$) fields. These operators are of great phenomenological interest in the non-supersymmetric case, since they are employed in the calculation of certain transition amplitudes among bound states of particles and in the extraction of meson and baryon form factors. Correlation functions of such operators calculated in lattice SQCD therefore provide interesting probes of physical properties of the theory. A proper renormalization of these operators is most often indispensable for the extraction of results from the lattice. The main objective of this work is the calculation of the quantum corrections to a complete basis of ``ultra local'' bilinear currents, using both dimensional regularization and lattice regularization. We consider both flavor singlet and nonsinglet operators. 

Within the SQCD formulation we compute all 2-pt Green's functions of bilinear operators, made out of quark and squark fields. Our computations are performed to one loop and to lowest order in the lattice spacing, $a$; also, in order to avoid potential infrared singularities in Green's functions of squark bilinears, we have employed massive chiral supermultiplets throughout. Quantum corrections induce mixing of some of the bilinear operators which we study, both among themselves and with gluon and gluino bilinears having the same quantum numbers; we compute all the corresponding mixing coefficients, in different renormalization schemes. 

\subsection{Lattice Action}
\label{sec2.1}
In our lattice calculation, we extend Wilson's formulation of the QCD action, to encompass SUSY partner fields as well. In this standard discretization quarks, squarks and gluinos live on the lattice sites, and gluons live on the links of the lattice: $U_\mu (x) = e^{i g a T^{\alpha} u_\mu^\alpha (x+a\hat{\mu}/2)}$; $\alpha$ is a color index in the adjoint representation of the gauge group. This formulation leaves no SUSY generators intact, and it also breaks chiral symmetry; it thus represents a ``worst case'' scenario, which is worth investigating in order to address the complications \cite{Giedt} which will arise in numerical simulations of SUSY theories. In our ongoing investigation we plan to address also improved actions, so that we can check to what extent some of the SUSY breaking effects can be alleviated. For Wilson-type quarks ($\psi$) and gluinos ($\lambda$), the Euclidean action ${\cal S}^{L}_{\rm SQCD}$ on the lattice becomes ($A_\pm$ are the squark field components):       
\bea
{\cal S}^{L}_{\rm SQCD} & = & a^4 \sum_x \Big[ \frac{N_c}{g^2} \sum_{\mu,\,\nu}\left(1-\frac{1}{N_c}\, {\rm Tr} U_{\mu\,\nu} \right ) + \sum_{\mu} {\rm Tr} \left(\bar \lambda_M \gamma_\mu {\cal{D}}_\mu\lambda_M \right ) - a \frac{r}{2} {\rm Tr}\left(\bar \lambda_M  {\cal{D}}^2 \lambda_M \right) \nonumber \\ 
&+&\sum_{\mu}\left( {\cal{D}}_\mu A_+^{\dagger}{\cal{D}}_\mu A_+ + {\cal{D}}_\mu A_- {\cal{D}}_\mu A_-^{\dagger}+ \bar \psi_D \gamma_\mu {\cal{D}}_\mu \psi_D \right) - a \frac{r}{2} \bar \psi_D  {\cal{D}}^2 \psi_D \nonumber \\
&+&i \sqrt2 g \big( A^{\dagger}_+ \bar{\lambda}^{\alpha}_M T^{\alpha} P_+ \psi_D  -  \bar{\psi}_D P_- \lambda^{\alpha}_M  T^{\alpha} A_+ +  A_- \bar{\lambda}^{\alpha}_M T^{\alpha} P_- \psi_D  -  \bar{\psi}_D P_+ \lambda^{\alpha}_M  T^{\alpha} A_-^{\dagger}\big)\nonumber\\  
&+& \frac{1}{2} g^2 (A^{\dagger}_+ T^{\alpha} A_+ -  A_- T^{\alpha} A^{\dagger}_-)^2 - m ( \bar \psi_D \psi_D - m A^{\dagger}_+ A_+  - m A_- A^{\dagger}_-)\Big] \,,
\label{susylagrLattice}
\eea
where: $U_{\mu \nu}(x) =U_\mu(x)U_\nu(x+a\hat\mu)U^\dagger_\mu(x+a\hat\nu)U_\nu^\dagger(x)$, and a summation over flavors is understood in the last three lines of Eq.~(\ref{susylagrLattice}) (the penultimate term in parentheses implies a summation over two independent flavor indices). The 4-vector $x$ is restricted to the values $x = na$, with $n$ being an integer 4-vector. The terms proportional to the Wilson parameter, $r$, eliminate the problem of fermion doubling, at the expense of breaking chiral invariance. In the limit $a \to 0$ the lattice action reproduces the continuum one.

A gauge-fixing term, together with the compensating ghost field term, must be added to the action, in order to avoid divergences from the  integration over gauge orbits; these terms are the same as in the non-supersymmetric case. Similarly, a standard ``measure'' term must be added to the action, in order to accound for the Jacobian in the change of integration variables: $U_\mu \to u_\mu$\,. All the details and definitions of the continuum and the lattice actions can be found in Ref.\cite{MC:2018}.

\subsection{Bilinear operators and their mixing}
\label{sec2.2}

In studying the properties of physical states, the main observables are Green's functions of operators made of quark fields, having the form ${\cal O}_i^\psi(x) = \bar \psi(x) \Gamma_i \psi(x)$, where $\Gamma_i$ denotes all possible distinct products of Dirac matrices, as well as operators made of squark fields ${\cal O}^A(x) = A^\dagger(x) A(x)$, along with operators of higher dimensionality. The matter fields are considered to be massive; in this way, we have control over infrared (IR) divergences. Ultraviolet (UV) divergences are treated by a standard regularization, either the lattice (L) or dimensional regularization (DR).

By investigating the behavior of dimension-2 and -3 bilinear operators under parity, $\cal{P}$, charge conjugation, $\cal{C}$, and exchange between the flavors of the two fields, $\cal{F}$, for the mass-degenerate case, we group them to the categories $S$, $P$, $V$, $A$ and $T$ in Tables~\ref{tb:non-singlet} and \ref{tb:singlet}, based on the trasformation properties of the quark bilinears. 
[Mixing with further operators, such as $A_+^{\dagger} A_+-A_- A_-^{\dagger}$ and $A_+^{\dagger} D_\mu A_-^{\dagger} - A_- D_\mu A_+$, is not allowed in the mass degenerate case, due to incompatible eigenvalues under $\cal{C} \times \cal{F} $; mixing is not allowed also in the non-degenerate case, since the mixing coefficients, being mass-independent, would necessarily coincide with those in the degenerate case which vanish by the above argument.]

\begin{table}[ht!]
\centering
\begin{tabular}{c|c|c|c}
\hline
\hline
\textbf{Operators}  & \,\,\,\,\textbf{$\cal{P}$}\,\,\,\, & \,\,\,\,\textbf{$\cal{C} \times \cal{F} $}\,\,\,\, &\textbf{Category}  \\ [0.5ex] \hline
$\bar \psi \psi$& $+$& $+$&$S$  \\[0.5ex]\hline
$\bar \psi \gamma_5 \psi$& $-$ &$+$ &$P$  \\[0.5ex]\hline
$\bar \psi \gamma_\mu \psi$&$(-1)^\mu$ & $-$ & $V$ \\[0.5ex]\hline
$\bar \psi \gamma_5 \gamma_\mu \psi$&$-(-1)^\mu$ &$+$&$AV$  \\[0.5ex]\hline
$\frac{1}{2}\bar \psi [\gamma_\mu , \gamma_\nu] \psi$ &$(-1)^\mu(-1)^\nu$ &$-$ &$T$  \\[0.5ex]\hline
$A_+^{\dagger} A_++A_- A_-^{\dagger}$&$+$&$+$&$S$  \\[0.5ex]\hline
$A_+^{\dagger} A_-^{\dagger} + A_- A_+$&$+$&$+$& $S$  \\[0.5ex]\hline
$A_+^{\dagger} A_-^{\dagger} - A_- A_+$&$-$&$+$& $P$ \\[0.5ex]\hline
$(m_f + m_{f'}) (A_+^{\dagger} A_+ + A_- A_-^{\dagger})$&$+$&$+$& $S$  \\[0.5ex]\hline
$(m_f - m_{f'}) (A_+^{\dagger} A_+ - A_- A_-^{\dagger})$&$-$&$+$& $P$  \\[0.5ex]\hline
$(m_f + m_{f'}) (A_+^{\dagger} A_-^{\dagger} + A_- A_+)$&$+$&$+$& $S$  \\[0.5ex]\hline
$(m_f + m_{f'}) (A_+^{\dagger} A_-^{\dagger} - A_- A_+)$&$-$&$+$& $P$  \\[0.5ex]\hline
$A_+^{\dagger} D_\mu A_+ + A_- D_\mu  A_-^{\dagger}$&$(-1)^\mu$&$-$& $V$  \\[0.5ex]\hline
$A_+^{\dagger} D_\mu A_+ - A_- D_\mu  A_-^{\dagger}$&$-(-1)^\mu$&$+$& $AV$  \\[0.5ex]\hline
$A_+^{\dagger} D_\mu A_-^{\dagger} + A_- D_\mu A_+$&$(-1)^\mu$&$-$&$V$ \\[0.5ex]\hline
\hline
\end{tabular}
\caption{Quark bilinears and other operators with which these can mix, in the flavor non-singlet case. Only gauge invariant operators appear in this case. Operators with lower dimensionality will mix on the lattice. All operators appearing in this table are eigenstates of $\cal{P}$ and $\cal{C} \times \cal{F}$. In the above operators, the matter fields should be explicitly identified with a flavor index. The flavor indices carried by the left fields ($f$) differ from those of right fields ($f'$). The shorthand $(-1)^\mu$ means $+$ for $\mu=0$ and $-$ for $\mu= 1, 2, 3$.}
\label{tb:non-singlet}
\end{table}

\begin{table}[ht!]
\centering
\begin{tabular}{ c| c} 
\hline
\hline
\textbf{Operators}  & \textbf{Category}  \\ [0.5ex] \hline
$\openone$& $S$  \\[0.5ex]\hline
${\rm Tr}\left(\bar \la \la\right)$& $S$  \\[0.5ex]\hline
${\rm Tr}\left(\bar \la \gamma_5 \la\right)$& $P$  \\[0.5ex]\hline
${\rm Tr}\left(\bar \la \gamma_\mu \la\right)$& $V$ \\[0.5ex]\hline
${\rm Tr}\left(\bar \la \gamma_5 \gamma_\mu \la\right)$& $AV$  \\[0.5ex]\hline
${\rm Tr}\left(\frac{1}{2}\bar \la [\gamma_\mu , \gamma_\nu] \la\right)$ & $T$  \\[0.5ex]\hline
\hline
\end{tabular}
\caption{Additional operators which can mix with quark bilinears in the flavor singlet case.}
\label{tb:singlet}
\end{table}

\begin{figure}[ht!]
\centering
\includegraphics[scale=0.30]{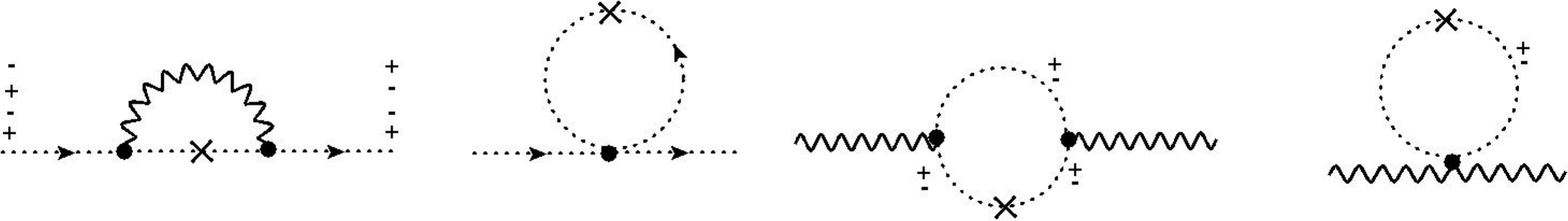}
\caption{One-loop Feynman diagrams leading to the renormalization of dimension-2 squark bilinear operators and to the potential mixing coefficients among themselves and/or (in the flavor singlet case) with gluon bilinears. A cross corresponds to squark operators. A wavy (dotted) line represents gluons (squarks). Squark lines are further marked with a $+$($-$) sign, to denote an $A_+ \, (A_-)$ field. A squark line arrow entering (exiting) a vertex denotes a $A_+$ ($A_+^{\dagger}$) field; the opposite is true for $A_-$ ($A_-^{\dagger}$) fields.
}
\label{mixing1}
\end{figure}

\begin{figure}[ht!]
\centering
\includegraphics[scale=0.50]{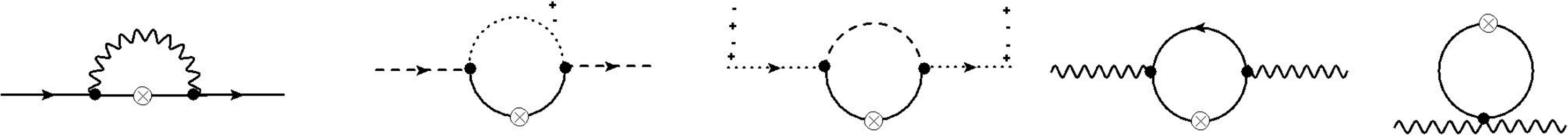}
\caption{One-loop Feynman diagrams leading to the renormalization of dimension-3 quark bilinear operators and to the potential mixing coefficients with gluino, squark and 
gluon bilinears. A circled cross corresponds to quark operators. A wavy (solid) line represents gluons (quarks). A dotted (dashed) line corresponds to squarks (gluinos). Squark lines are further marked with a $+$($-$) sign, to denote an $A_+ \, (A_-)$ field. A squark line arrow entering (exiting) a vertex denotes a $A_+$ ($A_+^{\dagger}$) field; the opposite is true for $A_-$ ($A_-^{\dagger}$) fields. Only the first and third diagram above are relevant to flavor nonsinglet operators.}
\label{mixing2}
\end{figure}

In order to calculate the one-loop mixing coefficients relevant to the squark- and quark-bilinear operators of lowest dimensionality, we must evaluate Feynman diagrams as shown in Figs.~\ref{mixing1} and \ref{mixing2}, respectively. In Fig.~\ref{mixing1} we have also included diagrams with external gluons; actually, since there are no BRST-invariant dimension-2 gluon operators, no mixing is expected to appear in this case, and we use this fact as a check on our perturbative results on the lattice. The diagrams in Fig.~\ref{mixing2} lead to the renormalization of dimension-3 quark bilinear operators and to the potential mixing coefficients with gluino, squark and gluon bilinears.

\newpage
\section{Details of the calculation and results}
\label{sec3}

The purpose of this section is to properly renormalize the quark and squark bilinear operators. A complete presentation of our results is relegated to a forthcoming publication~\cite{MC:2019}.
The definitions of the renormalization factors for matter fields and masses are (see Ref.~\cite{MC:2018} for details):
\bea
\psi^R &=& \sqrt{Z_\psi}\,\psi^B,\\
\label{condS}
A^R &=& \sqrt{Z_{A_\pm}}\,A^B, \\
m_\psi^R &=& Z_{{m}_\psi}\,m_\psi^B, \\ 
{{m^R_{A}}}^\dagger m_A^R &=& Z_{{m}_A}^\dagger {{m^B_{A}}}^\dagger m_A^B Z_{{m}_A},
\label{condMS}
\eea
where $B$ stands for the bare and $R$ for renormalized quantities and $Z_A$, $Z_{{m}_A}$ are $2\times2$ matrices corresponding to the doublet A.
Firstly, we calculate the case of squark bilinear operators in the continuum, where we regularize the theory in $D$ dimensions ($D=4-2\,\epsilon$). The squark bilinears,  ${\cal O}^A$, having dimension 2 can mix in principle with other dimension-2 operators. This entails studying the 2-pt Green's functions of ${\cal O}^A$ with external squark and gluon fields.  
There are four squark operators which we denote as $O^A_{\pm\,\pm}$ :
\bea
{\cal O}_{+\,+}^A(x) &=& A_+^{\dagger}(x)\, A_+(x) \\
{\cal O}_{+\,-}^A(x) &=& A_+^{\dagger}(x)\, A_-^{\dagger}(x)\\
{\cal O}_{-\,+}^A(x) &=& A_-(x)\, A_+(x)\\
{\cal O}_{-\,-}^A(x) &=& A_-(x)\, A_-^{\dagger}(x)
\eea
Given that all the quantities which we set out to calculate are $x$-independent, we will often apply an integration (or a summation on the lattice) over $x$ for convenience.

In order to check that there is no mixing with other Lorentz scalar dimension-2 gluon operators, we calculate the diagrams of the squark bilinear operators with external gluons. By studying the corresponding Green's functions we find that indeed this case receives no mixing, and thus flavor singlet squark operators cannot mix with the gluon bilinear $u_\mu u^\mu$. In particular, there was a cancelation of the pole part in the last two diagrams of Fig.~\ref{mixing1} leading to the expected result. 
Indeed, the 2-pt Green's functions of these operators with external gluons turn out to be finite and  transverse. Since $u_\mu u^\mu$ is neither gauge invariant, nor a BRST variation of some other operator, nor does it vanish by the  equations of motion, the lattice regulator also does not allow mixing with this operator. We check the above, calculating the same quantities on the lattice. Our results coincide with those of the continuum. 

Further, we have calculated the Green's functions of squark bilinear operators with external squarks (first two diagrams in Fig.~\ref{mixing1}), both on the lattice and in the continuum regularizations. Their renormalization is explicitly denoted as:
\be
   {{\cal O}_{a\,b}^A}^R = Z_{a\,b\,a'\,b'} {{{\cal O}_{a'\,b'}^A}^B}
\ee
where the indices ${a, b, a', b'}$ can take the values $+$ and $-$. Requiring that the bare lattice Green's functions, upon renormalization, lead to the same expressions as the continuum ones, we arrive at the following lattice renormalization factors: 
\bea
Z_{+\,+\,+\,+}^{L,\MSbar}=Z_{-\,-\,-\,-}^{L,\MSbar}&=& 1 - \frac{g^2\,C_F}{16\,\pi^2} \left(12.5586 + 2 \log(a^2 \bar\mu^2) \right)\\
Z_{+\,-\,+\,-}^{L,\MSbar}=Z_{-\,+\,-\,+}^{L,\MSbar}&=& 1 - \frac{g^2\,C_F}{16\,\pi^2}\left(20.1462\right)
\eea
Note that all squark operators do not mix among themselves and the operators ${\cal O}_{+\,-}^A$ and ${\cal O}_{-\,+}^A$ receive no logarithmic corrections up to one loop. Furthermore, all other components of $Z_{a\,b\,a'\,b'}$ vanish.

We now turn to the Green's functions of the quark bilinear operators, ${\cal{O}}_i^\psi$. From the first diagram in Fig.~\ref{mixing2} we calculate the renormalization factors $Z_{i}$ of ${\cal{O}}_i^\psi$. The other diagrams contribute to the mixing coefficients $z^\la_{i}$, $z^{\pm\,\pm}_{i}$, $z^{\pm D\pm}_{i}$, $z^{m\pm \pm}_{i}$, $z^u_{i}$, with gluino, squark (involving zero or one derivatives, or one power of the mass) and gluon bilinear operators, respectively. The expressions relevant for the mixing of each quark bilinear assume the following forms:
\bea
\label{scalar}
{{\cal O}_S^\psi}^R &=& Z_S {{\cal O}_S^\psi}^B + z^\la_S {{\cal O}_S^\la}^B + z^{+\,+}_S ({{\cal O}_{+\,+}^A}^B+{{\cal O}_{-\,-}^A}^B) + z^{+\,-}_S ({{\cal O}_{+\,-}^A}^B + {{\cal O}_{-\,+}^A}^B)  \\ \nonumber
&& + z^{m+\,+}_S (m_f + m_{f'})({{\cal O}_{+\,+}^A}^B+{{\cal O}_{-\,-}^A}^B) + z^{m+\,-}_S (m_f + m_{f'}) ({{\cal O}_{+\,-}^A}^B + {{\cal O}_{-\,+}^A}^B)   \\
{{\cal O}_P^\psi}^R &=& Z_P {{\cal O}_P^\psi}^B + z^\la_P {{\cal O}_P^\la}^B + 
z^{+\,-}_P ({{\cal O}_{+\,-}^A}^B - {{\cal O}_{-\,+}^A}^B) 
+z^{m+\,+}_P (m_f - m_{f'}) ({{\cal O}_{+\,+}^A}^B-{{\cal O}_{-\,-}^A}^B)\\ \nonumber
&& +z^{m+\,-}_P (m_f + m_{f'}) ({{\cal O}_{+\,-}^A}^B - {{\cal O}_{-\,+}^A}^B)  \\
{{\cal O}_{V,\mu}^\psi}^R &=& Z_{V} {{\cal O}_{V,\mu}^\psi}^B + z^\la_V {{\cal O}_{V,\mu}^\la}^B + z^{+D+}_V (A_+^{\dagger} D_\mu A_+ + A_- D_\mu  A_-^{\dagger}) \\\nonumber
&&+ z_V^{+D-}(A_+^{\dagger} D_\mu A_-^{\dagger} + A_- D_\mu A_+) +z^u_V u_\mu \partial_\nu u_\nu
\\
{{\cal O}_{AV,\mu}^\psi}^R &=& Z_{AV} {{\cal O}_{AV,\mu}^\psi}^B + z^\la_{AV} {{\cal O}_{AV,\mu}^\la}^B + z_{AV}^{+D+}(A_+^{\dagger} D_\mu A_+^{\dagger} - A_- D_\mu A_-)+z^u_{AV} \epsilon_{\mu\,\nu\,\rho\,\sigma}u_\nu \partial_\rho u_\sigma\\
{{\cal O}_T^\psi}^R &=& Z_T {{\cal O}_T^\psi}^B + z^\la_T {{\cal O}_T^\la}^B
\label{tensor}
\eea
On the rhs of Eqs.~(\ref{scalar})-(\ref{tensor}) there appear all operators that can possibly mix with those on the lhs; the tree-level Green's functions of these mixing operators naturally show up in the results for the one-loop Green's functions of the quark operators, thus allowing us to deduce the corresponding mixing coefficients. In the case of flavor nonsinglet operators, $z^\la_i$ and $z^u_i$ automatically vanish. The axial vector quark operator yields an expression related to the axial anomaly. The latter stems from the last two diagrams of Fig.~\ref{mixing2}, which involves non-supersymmetric particles; thus, it must reproduce the equivalent result in QCD. The lattice discretization should give the correct axial anomaly term in the continuum limit. Lastly, the tensor quark operator can only mix with the gluino one. 

By combining the lattice expressions with the renormalized Green's functions calculated in the continuum, we find for the renormalization factors: 
\bea
{Z_S}^{L,\MSbar} &=& 1 + \frac{g^2\,C_F}{16\,\pi^2}  \left( -13.1105 + \log(a^2\bar\mu^2)\right)\\
{Z_P}^{L,\MSbar} &=& 1 + \frac{g^2\,C_F}{16\,\pi^2}  \left(-22.7536 + \log(a^2\bar\mu^2)\right)\\
{Z_{V}}^{L,\MSbar} &=& 1 + \frac{g^2\,C_F}{16\,\pi^2}  \left(-20.7759 -2 \log(a^2\bar\mu^2)\right)\\
{Z_{AV}}^{L,\MSbar} &=& 1 + \frac{g^2\,C_F}{16\,\pi^2}  \left(-15.9544 -2 \log(a^2\bar\mu^2)\right)\\
{Z_T}^{L,\MSbar} &=& 1 + \frac{g^2\,C_F}{16\,\pi^2} \left(-17.1762 -3 \log(a^2 \bar\mu^2) \right).
\eea
We note that these factors are all gauge independent, as they should be in the $\MSbar$ scheme. The remaining quantities which we need to compute on the lattice are the mixing coefficients. These can be easily determined by computing the Green's functions corresponding to the last four diagrams of Fig.~\ref{mixing2} in the lattice regularization.  For several of these Green's functions, their difference from the corresponding ($\MSbar$-renormalized) functions in $DR$ vanishes, in the limit $a \to 0$, with no additional lattice contributions. 
The lattice one-loop expressions for the mixing coefficients, and in the mass-degenerate case, are presented here. By comparing the bare lattice Green's functions with the corresponding renormalized ones, we immediately obtain \footnote{The coefficient $z_P^{m++}$ stems from the third diagram in Fig.~\ref{mixing2}, evaluated with non-degenerate masses, see Ref.~\cite{MC:2019}.}:
\bea
\hspace{-1cm} z^\la_S=z^\la_P=z^\la_T=0&,&\,z^\la_V=\frac{g^2}{16\,\pi^2} \Big (4.4839 - \log\left(a^2 \bar \mu^2 \right) \Big),\,z^\la_{AV}= \frac{g^2}{16\,\pi^2} \Big ( 5.7087 + \log\left(a^2 \bar \mu^2 \right)\Big) \\
\,z^{+\,+}_S&=& 
\frac{g^2\,C_F}{16\,\pi^2} \frac{1}{a} 23.8429 r,\, \,\,
z^{+\,-}_S= 
-\frac{g^2\,C_F}{16\,\pi^2} \frac{1}{a} 8.9274 r\\
z^{m+\,+}_S&=& 
\frac{g^2\,C_F}{16\,\pi^2} \left (26.4484 + 8 \log\left(a^2 \bar \mu^2 \right) \right),\,\,\, z^{m+\,-}_S= 
\frac{g^2\,C_F}{16\,\pi^2}3.639\\
z^{+\,-}_P&=&
-\frac{g^2\,C_F}{16\,\pi^2} \frac{1}{a} 32.7704 r,\,\,\,
z^{m+\,-}_P=
-\frac{g^2\,C_F}{16\,\pi^2}14.8886\\
z^{+D+}_V&=&
-\frac{g^2\,C_F}{16\,\pi^2}\left(5.6888 + 8\log\left(a^2 \bar \mu^2 \right) \right),\,\,\,
z^{+D-}_V=
 -\frac{g^2\,C_F}{16\,\pi^2} 0.8693\\
z^{+D+}_{AV}&=&
\frac{g^2\,C_F}{16\,\pi^2} \left (14.6168 + 8\log\left(a^2 \bar \mu^2 \right) \right),\, \,\,z^{u}_V=z^{u}_{AV}=0
\eea
The above perturbative estimates of the renormalization factors $Z_i$ and of the mixing coefficients $z_i$ can be used for the determination of the properly renormalized ${\cal O}_i^\psi$ matrix elements.

\section{Summary of Results}
\label{summary}

In this work we have studied the mixing under renormalization for local bilinear operators in SQCD. We have calculated the one-loop renormalization factors and mixing coefficients for local quark operators and dimension-2 squark operators, both in Dimensional Regularization (DR) and on the Lattice (L), in the $\MSbar$ renormalization scheme. In the supersymmetric case more mixings arise as compared to QCD, due to the fact that the SQCD action contains more fields and interactions, indeed in QCD there appears no mixing in the Green's functions of local quark bilinears. As a prerequisite, we have computed the quark and squark inverse propagators and thus we have determined the multiplicative renormalization of these fields and of their masses, as well as the critical values of each mass $m_f$. One novel aspect of this work is that we use the SQCD action with nonzero masses $m_f$ throughout our computations, thus there emerge more mixing patterns among operators made of quark, squark, gluino and gluons. 

The renormalization factors and the operator mixings can be determined from the calculation of certain 2-pt Green's functions. We have calculated the Green's functions of the dimension-2 squark operators with external squarks and gluons. The latter shows that there is no mixing between squark and gluon operators and from the former we have extracted the renormalization factors for the dimension-2 squark operators. However, for the quark operators, which are dimension 3, there are more mixing patterns. We have calculated the Green's functions of the quark bilinears with external quarks, squarks, gluinos and gluons and we determined all renormalization factors and mixing cofficients to one loop.

\end{document}